\documentclass[preprint,12pt]{elsarticle}

\usepackage{amssymb}
\usepackage{amsmath}
\usepackage[matrix,arrow,curve]{xy}

\journal{Journal of Geometry and Physics}

\def\rd{\mathrm{ d}}
\def\g{\mathfrak{g}}

\newcommand {\subplus}{\mathop{{\subset}\llap{\raise 0.5pt\hbox{\normalfont\small+}\hskip 0.5pt}}}

\newtheorem{theo}{Theorem}[section]

\newtheorem{prop}{Proposition}[section]
\newtheorem{rem}{Remark}[section]
\newtheorem{dfn}{Definition}[section]
\newtheorem{ex}{Example}[section]

\makeatletter
\def\ps@pprintTitle{%
 \let\@oddhead\@empty
 \let\@evenhead\@empty
 \def\@oddfoot{}%
 \let\@evenfoot\@oddfoot}
\makeatother

\begin{document}

\begin{frontmatter}



\title{Graded geometry in gauge theories and beyond}


\author{Vladimir Salnikov}
\ead{vladimir.salnikov@unicaen.fr}
\ead[url]{www.vladimir-salnikov.org}

\address{Nicolas Oresme Mathematics Laboratory, University of Caen Lower Normandy,\newline 
  CS 14032, Bd. Mar\'echal Juin,  BP 5186,
  14032 Caen Cedex,
  France }

\begin{abstract}
We study some graded geometric constructions appearing naturally in the context of 
gauge theories. Inspired by a known relation of gauging with equivariant cohomology 
we generalize the latter notion to the case of arbitrary $Q$-manifolds introducing 
thus the concept of equivariant $Q$-cohomology.
Using this concept we describe a procedure for analysis of
gauge symmetries of given functionals as well as for constructing 
functionals (sigma models) invariant under an action of some gauge group.

As the main example of application of these constructions we consider 
the twisted Poisson sigma model.
We obtain it by a gauging-type procedure of the action of an essentially
infinite dimensional group 
and describe its symmetries in terms of classical differential geometry. 

We comment on other possible applications of the described concept including 
the analysis of supersymmetric gauge theories and higher structures.
\end{abstract}


\end{frontmatter}


\newpage
   \section{Introduction/preliminaries}
In this paper we describe the possibilities offered by graded geometry 
for the analysis of physical theories and some objects of classical 
differential geometry. We introduce a powerful tool --
the concept of equivariant $Q$-cohomology 
which is a natural extension of the definition of standard equivariant
cohomology to $Q$-manifolds.

In the first part of the paper after introducing the 
problem of gauging we briefly sketch some facts from 
the theory of $Q$-manifolds and fix the notations in the examples that are
important in what follows. 
In the section \ref{sec:eqQcoh} we define the notion of equivariant $Q$-cohomology
and explain, following the scheme of A.~Kotov and T.~Strobl (\cite{Kotov-Strobl}, \cite{GeneralizingGeometry}),
its relation to gauge invariance. Within this framework we recover 
explicitly the result of J.M.~Figueroa-O'Farrill and S.~Stanciu 
\cite{Stanciu} on the obstruction to gauging of the Wess-Zumino terms --
this shows also how ordinary equivariant cohomology can be obtained 
as a particular case of $Q$-cohomology. 
The section \ref{sec:HPSM} is entirely devoted to the analysis of the twisted 
Poisson sigma model (\cite{Klimcik-Strobl}): we describe the algebra of its symmetries 
and construct its functional using the procedure that we suggest as an alternative 
to standard gauging.
In the last section \ref{sec:courant} we give a purely mathematical application 
of the described concept, namely we propose a possible definition of equivariant 
cohomology for Courant algebroids. To conclude we also comment on other 
applications and some work in progress.

\subsection{Gauging problem,  Wess-Zumino terms}
The major part of this work is motivated by the gauging problem, which is important in 
theoretical physics. To give a simple example of the procedure 
consider $X \colon \Sigma^d \to M^n$ -- a map between two smooth manifolds 
of dimensions $d$ and $n$ respectively, and $B \in \Omega^d(M)$; 
in the physicist's terminology one would call $X$ a scalar field, $\Sigma$ -- world-sheet
and $M$ -- target.
Assume that a Lie group $G$ acts on $M$ and leaves $B$ invariant. This induces a $G$-action on  
$M^\Sigma$, which leaves invariant the functional $ S[X] = \int_\Sigma X^* B$. 
The invariance with respect to  $G$ is called a (global) \emph{rigid  invariance}.
The functional is called (locally) \emph{gauge invariant}, if it is invariant with respect 
to the group $G^\Sigma\equiv C^\infty(\Sigma,G)$; it is clear that 
in its original form $S[X]$ is not necessarily gauge invariant.

The procedure of gauging consists in modifying the functional $S$ in order 
to make it gauge invariant. This is usually  done by introducing new variables 
to $S$ controlling however that the result reduces to the initial 
functional when the additional variables are put to zero
-- these new variables may have some physical meaning in concrete applications. 
For the above functional the gauging problem can be solved 
by extending  $S$ to a functional $\widetilde{S}$ defined on 
$(X,A) \in M^\Sigma \times \Omega^1(\Sigma, \g), \g = Lie(G)$ 
by means of so-called minimal coupling.  For example if  $d=2$ the result is 
\begin{equation} \label{SBgb} \nonumber
 \widetilde{S}[X,A] = \int_\Sigma \left( X^* B - A^a  X^* \iota_{v_a} B + \frac{1}{2} A^a A^b  X^* \iota_{v_a} \iota_{v_b} B \right) , 
\end{equation}
where $A^a, A^b, \iota_{v_a}, \iota_{v_b}$ are defined by fixing the basis of $\g$.\footnote{Here and in the whole text the convention of summation over repeating indeces is adopted.}
Making again a remark about the physicist's terminology one would call
the variables $A$ the one-form valued (gauge) fields.

Suppose now that $\Sigma^d = \partial \Sigma^{d+1}$, $\tilde X \colon \Sigma^{d+1} \to M$ -- 
an extension of $X$ coinciding with it on $\Sigma^d$.
Let $H \in \Omega^{d+1}(M)$ be a closed differential form 
 invariant under the induced action of $G$. The functional 
$ S[X] = \int_{\Sigma^{d+1}} \tilde X^* H $ is the simplest one containing the so-called
\emph{Wess-Zumino term} (the integration is performed over the bulk of the worldsheet manifold, 
\cite{Witten}).
In contrast to the previous example gauging of this functional can be obstructed. 
More precisely in \cite{Stanciu, etc} it has been shown that
gauging is possible, if and only if $H$ permits an equivariantly closed extension. 
There is however a serious limitation for the procedure suggested in \cite{Stanciu}, 
namely the number of introduced gauge fields is equal to the dimension 
of the group $G$ making it not very practical for essentially infinite 
dimensional groups which do appear in applications. 
In what follows we will see how this issue can be treated 
within the framework of $Q$-bundles and in particular observe the 
situations when the extension of the target is not given explicitly by the 
Lie algebra of the group acting.

\subsection{Graded geometry, $Q$-manifolds} \label{sec:qman}
We are certainly not going to give a full introduction to graded geometry,
referring to nice sources like \cite{leites, bern_sem, Voronov, Roytenberg2002}.
Let us however give a definition of a $Q$-manifold and several examples of it. 
\begin{dfn}
A \emph{$Q$-manifold} is {graded manifold} equipped with a \emph{$Q$-structure}
-- a degree $1$ vector field $Q$ satisfying $[Q, Q] \equiv 2 Q^2 = 0$. 
\end{dfn}

\begin{ex} \label{ex:1}
  Consider any smooth manifold $M$, declare fiber linear coordinates 
  of the tangent bundle to it to be of degree $1$. The graded manifold 
  obtained like this is generally denoted $T[1]M$. It is equipped with 
  the de Rham differential that written in local coordinates ($x^i, \theta^i = \rd x^i$) 
  has the form  $\rd_{\text{dR}} = \theta^i \frac{\partial}{\partial x^i}$, 
  it thus can be viewed as a degree $1$ vector field squaring to zero.
\end{ex}

\begin{ex} \label{ex:2}
 Consider a Lie algebra $\g$, choose a basis of it and declare local coordinates $\xi^a$
 to be of degree 1. This graded manifold denoted $\g[1]$ is equipped with the 
 Chevalley-Eilenberg differential 
 $Q_{CE} = C^a_{bc} \xi^b \xi^c \frac{\partial}{\partial \xi^a}$, where $C^a_{bc}$
 are the structure constants of $\g$. $Q_{CE}^2 = 0$ is equivalent to the Jacobi identity.
\end{ex}

A more involved example is provided by a twisted Poisson manifold $M$. 
Let us recall that given a closed differential form $H \in \Omega^3(M)$ 
an almost Poisson bivector $\Pi$ is called \emph{twisted Poisson} if and only if 
it satisfies the twisted version of the Jacobi identity: 
$[\Pi, \Pi]_{SN} = (\Pi^\#)^{\otimes 3}(H)$, where 
$[\cdot; \cdot]_{SN}$ is the Schouten-Nijenhuis bracket of multivector fields 
and the right hand side of the equality denotes the full contraction of $H$
with $\Pi$. In this case the couple $(\Pi, H)$ is called a \emph{twisted Poisson structure}.

\begin{ex} \label{ex:3}
  Consider a cotangent bundle to a manifold $M$ equipped with a twisted Poisson 
  structure $(\Pi, H)$. A graded manifold obtained by shifting the 
  grading of the fiber linear coordinates $p_i$ by $1$ is usually denoted 
  $T^*[1]M$. For $C^{jk}_i(x) = \frac{\partial \pi^{jk}}{\partial x^i} + H_{ij'k'}\pi^{jj'}\pi^{kk'}$
  consider the degree $1$ vector field 
  $$
  Q_{\pi, H} = 
   \pi^{ij}p_{j} \frac{\partial}{\partial x^i} - 
      \frac{1}{2} C_{i}^{jk}p_{j}p_{k} \frac{\partial}{\partial p_i}. 
  $$
  $Q_{\pi, H}^2 = 0$ is precisely equivalent to the twisted Jacobi identity. 
  Note that in the ``untwisted'' case of $H = 0$ this vector field is hamiltonian with respect 
  to the canonical Poisson structure on $T^*[1]M$:
  $Q_{\pi, 0} = \{\frac{1}{2}\pi^{ij}p_i p_j, \cdot \}$.
\end{ex}

This is certainly a non-exhaustive list of examples: the objects like 
Dirac structures, Lie algebroids (cf. \cite{Vaintrob}), Courant algebroids also can fit to the picture. We will discuss the latter one in more details in section \ref{sec:courant}.

Having defined the $Q$-structure on a graded manifold one has also a natural definition 
of a morphism:
\begin{dfn}
  Given two $Q$-manifolds (${\cal M}_1, Q_1$), (${\cal M}_2, Q_2$),
  a degree preserving map $f\colon {\cal M}_1 \to {\cal M}_2$, is a \emph{$Q$-morphism} if and only if
  on all superfunctions it commutes with the action of the respective vector fields:
  $Q_1 f^* - f^* Q_2  = 0$.  
\end{dfn}

A generic degree preserving map $\varphi$ fails to be a $Q$-morphism but there is always a canonical 
construction permitting to extend $\varphi$ to a $Q$-morphism. More precisely the following 
statement holds true.
\begin{prop}\label{prop:q-morphism}
Given a degree preserving 
 map $\varphi$ between $Q$-manifolds (${\cal M}_1, Q_1$) and 
  $({\cal M}_2, Q_2)$, there exists a $Q$-morphism $f$ between  the $Q$-manifolds (${\cal M}_1, Q_1$) and 
  $(\tilde {\cal M}, \tilde Q) = (T[1]{\cal M}, \rd _{DR} + {\cal L}_{Q_2})$, covering $\varphi$.
\end{prop}
\textbf{Proof.} Fix a coordinate system $q^a$ on ${\cal M}_2$, the vector field $Q_2$ reads 
$Q^a \frac{\partial}{\partial q^a}$. From the Cartan's magic formula one immediately sees that 
$\tilde Q q^a = \rd q^a + Q^a$. Since we want $f$ to cover $\varphi$, $f^*$ is forced to coincide 
with $\varphi$ on 
all the functions of $q$. To extend it to a $Q$-morphism on the whole $\tilde {\cal M}$ it 
is sufficient to define $f^*(\rd q^a) = Q_1 f^*(q^a) - f^*(Q^a)$ and note that $\tilde Q$ commutes with $\rd$. 
For details and a more geometric interpretation, 
see the discussion on the ``field strength'' in \cite{Kotov-Strobl} as well as the 
proposition 3.3 there. 
$\square$

\begin{rem}
 It is important to note that in this construction there is a double 
 grading appearing naturally: each homogeneous function on the resulting graded 
 manifold $\tilde {\cal M}$ can inherit a degree from ${\cal M}_2$ or from 
 the shift in the tangent bundle. We will use the Bernstein--Leites sign convention 
 for treating this double grading, i.e. the sign in the commutation relations is governed 
 by the sum of the degrees of each element.  
\end{rem}

\section{Equivariant $Q$-cohomology } \label{sec:eqQcoh}
We have mentioned in the previous section that the problem of gauging of the Wess-Zumino terms 
is related to equivariant extensions. 
Just to motivate the coming definitions let us note that in the ordinary case 
main objects for computing equivariant cohomology are the contractions of differential forms 
with vector fields induced by the group action and the de Rham differential
(for details see for example \cite{Alekseev, yellow}). In the language of graded manifolds 
the contraction decreases the degree of any differential form (viewed as a superfunction) 
and the de Rham differential increases. Let us now generalize this construction to 
arbitrary $Q$-manifolds.

Let $({\cal M}, Q)$ be a $Q$-manifold, and let 
${\cal G}$ be a subalgebra of degree $-1$ 
vector fields $\varepsilon$ on ${\cal M}$, closed with respect to the $Q$-derived bracket 
($[\varepsilon, \varepsilon']_Q \equiv [\varepsilon, [Q, \varepsilon']]$).
We consider the action of $Q$ on superfunctions on ${\cal M}$ 
as a \emph{generalized differential} and the action 
of any $\varepsilon$ 
as a \emph{generalized contraction}.
\begin{dfn} \label{def:Ghor}
A superfunction $\omega$ on $ {\cal M}$ is
\emph{${\cal G}$-horizontal} if and only if \newline
$ \varepsilon \omega = 0$, $\forall \varepsilon \in {\cal G}$.
\end{dfn}

\begin{dfn}  \label{def:Geq}
A superfunction $\omega$ on $ {\cal M}$ is 
\emph{${\cal G}$-equivariant} if and only if $(ad_{Q} \varepsilon) \omega \equiv 
[Q, \varepsilon] \omega = 0$, $\forall \varepsilon \in {\cal G}$.
\end{dfn}

\begin{dfn}  \label{def:Gbasic}
 We call a superfunction  $\omega$ on $ {\cal M}$ 
\emph{${\cal G}$-basic} if and only if it is ${\cal G}$-horizontal and ${\cal G}$-equivariant.
\end{dfn}

\begin{rem}
  Like in the standard case within the framework of the above definitions 
  one can consider a family of equivariant differentials 
  $\rd_{\varepsilon} = Q + \varepsilon$, each of them does not square to zero 
  by itself, but does on ${\cal G}$-basic superfunctions.
\end{rem}
\begin{rem} \label{rem:Gbasic}
 It is easy to see that for $Q$-closed superfunctions being ${\cal G}$-horizontal is equivalent 
 to being ${\cal G}$-basic. 
\end{rem}
\begin{rem}
 The $Q$-derived bracket need not be skew-symmetric, it defines thus a Loday- 
 but in general not a Lie algebra structure.
 However, the definitions above are also valid for the particular case when 
 ${\cal G}$ is isomorphic to a Lie algebra ${\g} = Lie(G)$, like in the classical 
 definition of equivariant cohomology. In this case the restriction of the $Q$-derived
 bracket to the considered vector fields is necessarily skew-symmetric. 
\end{rem}

\subsection{Gauge invariance and $Q$-bundles } \label{sec:q-bundles}
The reason for considering these objects is that they appear naturally in the 
description of sigma models via $Q$-bundles proposed by A.~Kotov and 
T.~Strobl (\cite{Kotov-Strobl}). Let us sketch the approach here. 
 
Associate  $Q$-manifolds $({\cal M}_1, Q_1)$ 
and $({\cal M}_2, Q_2)$ respectively to the world-sheet and the target of a gauge theory
and encode the fields in a degree preserving map $\varphi$ between them. 
As we have seen in the proposition \ref{prop:q-morphism}, one can 
lift $\varphi$ to a $Q$-morphism. One should actually extend the target even more 
by considering the direct product $ (\hat {\cal M}, \hat Q) = 
({\cal M}_1 \times \tilde{\cal M}, Q_1 + \tilde Q)$ and the appropriate extension of 
$\varphi$ and $f$, this gives a $Q$-bundle -- see the diagram (\ref{diag}) where by abuse of 
notation we call the extended maps by the same letters $\varphi$ and $f$.
\begin{equation} \label{diag} 
\xymatrix{
      && ( {\cal M}_1 \times \tilde {\cal M}) \circlearrowleft \hat Q \ar@/^/ [ddll]^{pr_1}   \\
  \\
 {Q_1 \circlearrowright\cal M}_1 \ar@/^/[uurr]^{f} \ar [rr]^{\varphi}&& {\cal M}_1 \times {\cal M}_2 \ar [uu]& 
} 
\end{equation}
The key idea is that within the framework of this construction the 
gauge transformations can be parametrized by 
$\hat \varepsilon$ -- vector fields on ${\cal M}_1 \times \tilde {\cal M}$
of total degree $-1$, vertical with respect to the projection $pr_1$ to ${\cal M}_1$:
\begin{equation} \label{gt}
 \delta_{\varepsilon}  ( f^*\cdot) = 
   f^*(  [\hat Q, \hat \varepsilon] \cdot).   
\end{equation}
One can consider separately the dependence of $\hat \varepsilon$
on ${\cal M}_1$ and $\tilde {\cal M}$, and thus construct the algebra ${\cal G}$
of degree $-1$ vector fields on $\tilde {\cal M}$.
For the functionals of the form $S = \int_{\Sigma^{d+1}} f^*(\omega)$ gauge invariance 
would be guaranteed precisely by the condition of $\omega$ being ${\cal G}$-basic (definition 
\ref{def:Gbasic}) and $\tilde Q$-closed (cf. also the remark \ref{rem:Gbasic}).

\begin{rem}
  A natural question to ask is how generic is the situation when one can proceed with the 
  above construction. For the world-sheet manifold usually there is no problem: one often considers 
  $T[1]\Sigma$ as the $Q$-manifold (cf. example \ref{ex:1}). For the target
  according to \cite{Melchior-Thomas} the $Q$-structure exists when field equations satisfy 
  a certain type of Bianchi identities, which is not a very restrictive condition. 
\end{rem}

\subsection{ Gauging via equivariant $Q$-cohomology }
In view of the previous subsection it is natural to consider the integrand of the 
functional with the rigid symmetry group in the form of a pull-back by a $Q$-morphism 
from the target manifold of some superfunction $\omega$. 
The gauging problem reduces to finding a ${\cal G}$-basic 
 $\tilde Q$-closed extension $\tilde \omega$ of this superfunction.
As we have understood, the necessary data for this procedure 
is the geometry of the target encoded in the $Q$-structure and 
the morphism from the algebra of symmetries to the algebra ${\cal G}$ of degree $-1$ vector fields with a 
$\tilde Q$-derived bracket. The rest is a straightforward application 
of the conditions coming from the definition \ref{def:Gbasic}:
\begin{equation} \label{gauging}
  \tilde Q \tilde \omega = 0, \qquad \tilde \varepsilon \tilde \omega = 0, \quad
  \forall \tilde \varepsilon \in {\cal G}.
\end{equation}

As an example of this procedure let us consider gauging of the Wess-Zumino term 
for $dim \Sigma = 2$ and the extension of the target being governed by a Lie group $G$
acting on $M$. 
An appropriate geometric structure for this case is the action Lie algebroid 
$E = M \times \g$. Declaring like in the example \ref{ex:2} the coordinates 
on $\g$ to be of degree $1$ we obtain the graded manifold usually denoted 
as $E[1]$ with a $Q$-structure $Q_g = \rho - Q_{CE}$, where $\rho \colon \g \to \mathfrak{X}(M)$ is 
the action of $\g$ on $M$. In some local chart 
$Q_g = \xi^a\rho_a^i\frac{\partial}{\partial x^i} - C^a_{bc} \xi^b \xi^c \frac{\partial}{\partial \xi^a}$.
As explained above, we consider the target $Q$-manifold $(\tilde {\cal M}, \tilde Q) = (T[1]E[1], \rd + 
{\cal L}_{Q_g})$. For the symmetry algebra the construction is also rather straightforward:
we consider the degree $-1$ vector fields on $E[1]$ that are all of the form 
$\varepsilon = \varepsilon^a \frac{\partial}{\partial \xi^a}$ and lift them to $T[1]E[1]$ by 
a Lie derivative $\tilde \varepsilon = {\cal L}_{\varepsilon}$. One can check by a straightforward
computation that the $\tilde Q$-derived bracket of two such vector fields is 
governed by a Lie-algebra bracket of the respective elements of $\g$. 
A $3$-form $H$ on $M$ can be viewed as a superfunction on $T[1]E[1]$, the problem 
of extending the Wess-Zumino term defined by it can thus be formulated in the $Q$-language 
presented above. 

To solve the extension problem let us apply the conditions (\ref{gauging})
to the most general degree superfunction $\tilde H$ on $\tilde {\cal M}$ writing it in 
local coordinates. It is convenient to choose the basis generated by $x^i, \xi^a, \tilde Q x^i, 
\tilde Q \xi^a$ of total degrees $0, 1, 1$ and $2$ respectively.
The first condition of $\tilde Q$-closedness reduces the extension to 
$
\tilde H = \frac{1}{6} H_{ijk} \tilde Qx^i \tilde Q x^j \tilde Q x^k + E_{ia,j}\tilde Qx^i \tilde Q x^j \xi^a +
 \frac{1}{2}F_{ab,i} \tilde Qx^i \xi^a \xi^b + E_{ia} \tilde Qx^i \tilde Q \xi^a + F_{ab} \xi^a \tilde Q \xi^b
$, subject to $F_{(ab)} = 0$.\footnote{Here and in the whole text $\cdot_{,i}$  denotes the  
derivative w.r.t. $x^i$; $_{(\cdot\cdot)}$ -- symmetrized and $_{[\cdot\cdot]}$ antisymmetrized indeces} 
The second condition amounts to several equalities, among which on has:
$$
  \frac{1}{2} H_{ijk} \varepsilon^a \rho_a^i + (E_{j]a}\varepsilon^a)_{,[k} = 0, \qquad
  (E_{ia}\rho^i_b + F_{ba})\varepsilon^b.
$$
They restrict the extension to 
$$
  \tilde H = \frac{1}{6} H_{ijk} \tilde Qx^i \tilde Q x^j \tilde Q x^k + 
  \tilde Q (\frac{1}{2}E_{ja}\rho^j_b \xi^a\xi^b - E_{ia}\tilde Qx^i \xi^a),  
$$
that repeats the solution given in \cite{Stanciu}. 

Remembering about the 
antisymmetry of $F_{ab} = E_{ia}\rho^i_b$ one recovers the same obstructions to gauging 
as in \cite{Stanciu}, that 
can be interpreted as the existence of the equivariantly closed extension of $H$
(in terms of standard equivariant de Rham theory). This observation permits 
to validate the definitions \ref{def:Ghor} - \ref{def:Gbasic} in the sense that 
they indeed generalize the standard picture. 

\begin{rem}
  Some similar results about equivariant $Q$-cohomology are discussed in \cite{mehta}, where it is shown 
 that, when the action is free and proper, then it agrees with the $Q$-cohomology of the quotient.
\end{rem}

\section{Twisted Poisson sigma model} \label{sec:HPSM}
As we have already mentioned, the approach of the previous section 
is not limited to the target extended by the group to be gauged. 
In this section we will describe in details the application of it to analysis 
of the twisted Poisson sigma model.

As in the previous example the world-sheet is given by a smooth manifold 
$\Sigma^2$ (closed, orientable, with no boundary, $dim = 2$). 
The target is a smooth manifold $M^n$ with a (twisted) Poisson structure $(\Pi, H)$
(cf. example \ref{ex:3}). The functional over the space 
 of vector bundle morphisms $T\Sigma \to T^*M$ reads
 \begin{equation} \label{HPSM}
    S[X,A] = \int_{\Sigma} A_i \wedge \rd X^i + \frac{1}{2}\pi^{ij}A_i \wedge A_j + \int_{\Sigma^3} H,
 \end{equation}
where $X^i : \Sigma \to M$ are scalar fields and $A_i \in \Omega^1(\Sigma, X^* T^* M)$ --
$1$-form valued (``vector'') fields. 
For details about the definition, motivation and physical 
significance of the model one can refer to original sources: \cite{Schaller-Strobl}, \cite{Klimcik-Strobl},
as well as \cite{Ikeda}, \cite{Park}, \cite{Severa-Weinstein}. 
One should also note that the model is related to 
derivation of the famous Kontsevitch's formula (\cite{Kontsevich}) for quantization of Poisson manifolds.

\subsection{Symmetries of twisted PSM}
As it is suggested by the examples in the beginning of the section \ref{sec:qman}
the natural $Q$-manifolds to use as worldsheet and target are $(T[1]\Sigma, d_{dR})$
and $(T^*[1]M, Q_{\Pi, H})$. To recover the $Q$-morphism from the proposition 
\ref{prop:q-morphism} we lift the picture to $(\tilde {\cal M}, \tilde Q) = (T[1]T^*[1]M, \rd + {\cal L}_{Q_{\Pi,H}})$.
The most general degree $-1$ vector field  
on $T^*[1]M$ reads $\varepsilon = \varepsilon_i \frac{\partial}{\partial p_i}$.
To get into the framework of the section \ref{sec:q-bundles} we lift this vector 
field to $\tilde {\cal M}$ by the Lie derivative. The $\tilde Q$-derived bracket of 
such vector fields induces on sections of $T^*M$ a Lie algebra structure 
(twisted Poisson Lie algebroid bracket):
\begin{equation} \label{eps_br}
 [\varepsilon^1, \varepsilon^2] = {\cal L}_{\pi^{\#} \varepsilon^1} \varepsilon^2 - 
{\cal L}_{\pi^{\#} \varepsilon^2}\varepsilon^1 - 
\rd (\pi(\varepsilon^1, \varepsilon^2)) + \iota_{\pi^{\#}\varepsilon^1} \iota_{\pi^{\#}\varepsilon^2} H,
\end{equation}
defining thus a Lie algebra that we denote ${\cal G}$. 
With this natural geometric construction we can already formulate 
a statement about gauge symmetries of the twisted PSM.
\begin{theo} \label{th:sym}
Any smooth map from $\Sigma$ to the space $\Gamma(T^*M)$ of sections 
of the cotangent bundle to $M$ defines an infinitesimal 
gauge transformation of the twisted PSM in the sense of equation (\ref{gt})
if and only if for any point $\sigma \in \Sigma$ the section $\varepsilon$
 satisfies 
 \begin{equation} \label{magic3}
  \rd \varepsilon + \iota_{\pi^\# \varepsilon}H = 0,
  \end{equation}
 where $\rd$ is the de Rham differential on $M$.
\end{theo}
\textbf{Proof} (Sketch)\textbf{.} To prove the statement one needs to rewrite 
the functional of the twisted PSM in the form $S = \int_{\Sigma^3} f^*(\tilde H)$
and apply to $\tilde H$ the conditions (\ref{gauging}). The first one will be satisfied 
automatically since $H$ is a closed form, the second one will reduce to the 
equation (\ref{magic3}). We will give an explicit formula in the 
proof of the theorem \ref{prop:reverse}. $\square$
\begin{rem}
 This theorem permits to construct all the essential gauge transformations of the twisted PSM (a generating set), 
 we will however return to the question of enlarging the algebra of symmetries.
 The theorem specializes to the ordinary ``untwisted'' PSM by setting $H = 0$
 and one recovers the (properly understood) condition on $\varepsilon$ from \cite{Bojowald-Kotov-Strobl}.
\end{rem}

If we consider the result of application of the equivariant $Q$-cohomology procedure 
for the subalgebra of ${\cal G}$ defined by (\ref{magic3}), it can be ambiguous. 
The way out is to consider the full (modulo some technical assumptions) 
algebra $\tilde {\cal G}$ of degree $-1$ vector fields on $\tilde {\cal M}$. 
To define $\tilde {\cal G}$ consider 
$\tilde \varepsilon$ -- a degree $-1$ vector field on $\tilde {\cal M}$
of the form 
\begin{equation} \label{eps}
\tilde \varepsilon = \varepsilon_i \frac{\partial}{\partial p_i} +
\alpha_{ij}\theta^j\frac{\partial}{\partial \psi_i}
\end{equation}

\begin{rem} \label{barred}
 Let us note, that the vector field $\tilde \varepsilon$ introduced above is not of the 
most general form that one could have on $\tilde {\cal M}$: compare it with
$$
  \tilde \varepsilon = \varepsilon_i \frac{\partial}{\partial p_i} + 
  \bar \varepsilon^i \frac{\partial}{\partial \theta^i} +
  (\alpha_{ij}\theta^j + \bar \alpha_i^j p_j)\frac{\partial}{\partial \psi_i}
$$
But the Lie bracket of two such vector fields would in general produce a degree $-2$ vector field, namely
$$
 [\tilde \varepsilon,\tilde \varepsilon'] = (\bar \alpha_{i}^{'j} \varepsilon_j - 
  \bar \alpha_i^j \varepsilon'_j +  \alpha'_{ij} \bar\varepsilon^j - 
  \alpha_{ij} \bar\varepsilon'^j) \frac{\partial}{\partial \psi_i}  
$$
So, to stay inside the set of vector fields of degree precisely $-1$ and generalize $\tilde \varepsilon = {\cal L}_{\varepsilon}$ we have to keep only the ``unbarred'' components.
\end{rem}
It turns out, that this Lie algebra is rather natural, i.e. one can describe it up to isomorphism
using classical differential geometry. More precisely the following proposition holds:
\begin{prop} \label{prop:alg}
  The Lie algebra $({\tilde {\cal G}}, [ , ]_{\tilde Q})$ is isomorphic to the semi-direct product of Lie 
algebras ${\cal G} \subplus  {\cal A}$, where 
 ${\cal A}$ is a Lie algebra of covariant 2-tensors on $M$ with a bracket given by
\begin{equation} \label{stup}
  [\bar \alpha,\bar \beta]
   =   <\pi^{23}, \bar\alpha\otimes\bar\beta - \bar\beta\otimes\bar\alpha >,
\end{equation}
(the upper indeces ``$23$'' of $\pi$ stand for the contraction on the $2$d and $3$rd entry of the tensor 
product);  
$\cal G$ acts on $\cal A$ by
\begin{equation}
 \rho(\varepsilon) (\bar \alpha) =   {\cal L}_{\pi^{\#}\varepsilon}(\bar \alpha) - 
 <\pi^{23}, {\cal D}_H \varepsilon\otimes  \bar \alpha  >, \quad
 \text{for }  {\cal D}_H \varepsilon= \rd\varepsilon + \iota_{ \pi^{\#} \varepsilon }H. 
\label{action}
\end{equation}
\end{prop}
\textbf{Proof.}
First we have to show that the semi-direct product ${\cal G} \subplus  {\cal A}$ is well defined, namely 
that the operation given by (\ref{action}) indeed defines an action in agreement with a somewhat artificial 
Lie bracket (\ref{stup}). 
One can prove this fact by direct computations in some local chart, but it is rather lengthy, so to simplify 
it we notice that the first Lie derivative part of (\ref{action}) is itself a Lie algebra action.
Secondly, one may notice that the operator ${\cal D}_H$, defined on $1$-forms
behaves nicely with respect to the bracket, i.e.
$$
  {\cal D}_H [\varepsilon^1, \varepsilon^2] = 
  {\cal L}_{ \pi^{\#} \varepsilon^1} ({\cal D}_H \varepsilon^2)
  - {\cal L}_{ \pi^{\#} \varepsilon^2} ({\cal D}_H \varepsilon^1).
$$
The remaining part of the proof is just application of these observations to compare the right- and 
left-hand-sides in the definition of a Lie algebra action. 

Now we will construct explicitly the maps defining the exact sequence
$$
  0 \to {\cal A} \to \tilde{\cal G} \to {\cal G} \to 0
$$
We can notice, that the first term in the formula (\ref{eps}) for the vector field 
$\tilde \varepsilon$  corresponds precisely to the element 
$\varepsilon \in {\cal G}$, 
from it we can recover the bracket of $1$-forms as the derived bracket of vector fields
($\frac{\partial}{\partial p_i}$ part).
If we perform a change of coordinates in (\ref{eps}) we notice that $\alpha_{ij}$ doesn't 
transform as a tensor, but  being 
 corrected by subtracting $\varepsilon_{i,j}$ it 
produces the $2$-tensor $\bar \alpha \in {\cal A}$ and the respective bracket.
This defines the desired isomorphism. $\square$
\begin{rem}
Let us also note, that in the computations of \cite{Kotov-Strobl} for the untwisted case 
the second term is defined by the lift of the vector field from $T^*M$ to $TT^*M$
by the Lie-derivative $\tilde \varepsilon = {\cal L}_{\varepsilon}$.  That is the Lie algebra
considered there is ${\cal G}$ itself. 
(We used a similar construction in the theorem \ref{th:sym}).
And this lift in fact 
gives the form of the correction necessary to recover the Lie-algebra ${\cal A}$ in our case.
\end{rem}

Now we define the 
condition that will be responsible for gauge invariance of the sigma model.
Let us consider a subalgebra ${\cal GT}\subset \tilde {\cal G}$ defined by 
\begin{equation}
  \rd\varepsilon + \iota_{ \pi^{\#} \varepsilon }H = 0, \qquad  \bar \alpha^A = 0 ,  
   \label{magic_big}
\end{equation}
where $\bar \alpha^A$ is the antisymmetrization of the tensor $\bar \alpha$.
Restricted to this subalgebra the Lie bracket (\ref{eps_br})
simplifies to 
$
  [\varepsilon^1, \varepsilon^2] = 
   \rd (\pi(\varepsilon^2, \varepsilon^1)) + \iota_{\pi^{\#}\varepsilon^2} \iota_{\pi^{\#}\varepsilon^1} H
$
and the action (\ref{action}) to 
$
 \rho(\varepsilon) (\bar \alpha) =   {\cal L}_{\pi^{\#}\varepsilon}(\bar \alpha),
$
which preserves the symmetric tensor property.
\begin{rem}
  To define gauge transformations from ${\cal GT}$ using (\ref{gt})
  the first equality in (\ref{magic_big}) is necessary since it just repeats 
  (\ref{magic3}) from the theorem \ref{th:sym}. 
  The second one is the description of the extension of the 
  algebra of symmetries by gauge transformations that are trivial 
  in the sense of \cite{Henneaux-Teitelboim}.
\end{rem}

\subsection{Twisted Poisson sigma model from gauging}
Having defined the algebra ${\cal GT}$ we are ready to formulate the ``converse''
statement to the theorem \ref{th:sym}.
Given a closed $3$-form $H$ on $M$ we can view it as a superfunction  on $\tilde {\cal M} = T[1]T^*[1]M$. 
We search for a ${\cal GT}$-equivariantly closed extension
(in the generalized sense of section \ref{sec:eqQcoh}) of $H$, that is 
a $3$-form (or better to say a degree $3$ superfunction) $\tilde H$ on $\tilde {\cal M}$ which is 
$\tilde Q$-closed, ${\cal GT}$-equivariant and starts with $H$. The following statement holds true:
\begin{theo} \label{prop:reverse} $\,$ \newline
  \textbf{1.} Consider the graded manifold ${\cal M} = T[1]T^*[1]M$, equipped 
  with the $Q$-structure $\tilde Q = \tilde Q_{\Pi,H}$, governed by an $H$-twisted 
  Poisson bivector $\Pi$, such that the pull-back of $H$ to a dense set of orbits of $\Pi$
  is not vanishing identically. 
  The ${\cal GT}$-equivariantly closed extension of a given $3$-form
  $H$ is determined uniquely by  
  $$
    \tilde H = \frac{1}{6}H_{ijk}\rd x^i \rd x^j \rd x^k + 
    \frac{1}{2}H_{ijk}\pi^{k'k}\rd x^i \rd x^j p_{k'} + \rd p_i \rd x^i. 
  $$
 \newline
\textbf{2.}  Being pulled-back by $f^*$ this extension defines the integrand of the twisted Poisson 
   sigma model functional.
\end{theo}
\textbf{Proof.} 
Although the natural local coordinates one 
introduces on the fiber $\tilde {\cal M}$ of 
the extended target  are $(x^i (0), p_i (1), \rd x^i (1), \rd p_i (2))$\footnote{The numbers 
in brackets denote the total degree of each coordinate.}
it is more convenient to perform the computations in the basis generated by $\tilde Q$ 
as a differential, i.e. $(x^i (0), p_i (1), \tilde Q x^i(1), \tilde Q p_i(2))$
(see also the proposition \ref{prop:q-morphism}). The following 
expression for the vector field $\tilde Q$ makes this coordinate change explicit:
\begin{eqnarray} 
  \tilde Q = \left(\rd x^i + \pi^{i'i}p_{i'}\right)\frac{\partial}{\partial x^i} + 
   \left(\rd p_i - \frac{1}{2}C_i^{jk}p_jp_k\right)\frac{\partial}{\partial p_i} + 
   \nonumber\\ \label{Q} 
   + \left(\rd(\pi^{i'i}p_{i'})\right)\frac{\partial}{\partial \rd x^i} 
   +\left( \rd (-\frac{1}{2}C_i^{jk}p_jp_k)\right)\frac{\partial}{\partial \rd p_i} 
\end{eqnarray}
The most general degree $3$ superfunction on $\tilde {\cal M}$ reads
\begin{eqnarray}
  \omega = A_{ijk} \tilde Q x^i \tilde Q x^j \tilde Q x^k &+& B_{ij}^k\tilde Q x^i\tilde Q x^jp_k \nonumber \\
  + K_i^{jk}\tilde Q x^i p_j p_k &+& D^{ijk}p_i p_j p_k + E^j_i \tilde Q x^i \tilde Q p_j + F^{ij} \tilde Q p_i p_j,
\nonumber
\end{eqnarray}
where the coefficients are superfunctions of degree $0$, i.e. they depend only on $x$. 
On this superfunction we impose the condition of being ${\cal GT}$-basic in the sense 
of definition \ref{def:Gbasic}.
The condition of $\tilde Q$-closedness reduces $\omega$ to 
$$
  \omega' = A_{ijk} \tilde Q x^i \tilde Q x^j \tilde Q x^k + E_{i,j}^k\tilde Q x^i\tilde Q x^jp_k
  + \frac{1}{2}F_{,i}^{jk}\tilde Q x^i p_j p_k + E^j_i \tilde Q x^i \tilde Q p_j + F^{ij} \tilde Q p_i p_j,
$$
for $A_{[ijk,l]} = 0$ and $F^{(ij)} = 0$. 

Let us compute the action of the vector field $\tilde \varepsilon$
on this superfunction. Using its value on the 
generators: $\tilde \varepsilon \tilde Q x^i = \pi^{i'i}\varepsilon_{i'}, \quad 
\tilde \varepsilon p_i =  \varepsilon_i$, and \newline $\tilde \varepsilon \tilde Q p_i = 
(\alpha_{ij} + \varepsilon_{i,j})\tilde Q x^j -  ( (\alpha_{ij} + \varepsilon_{i,j})\pi^{j'j}  
- C^{jk}_{i}\varepsilon_j ) p_k,$
we obtain four types of terms proportional to $\tilde Qx^j\tilde Qx^k, \, \tilde Qx^jp_k, \,
p_jp_k, \, \tilde Qp_i$. The fact that they should vanish for all
$\tilde \varepsilon \in {\cal GT}$ gives respectively  
the following conditions:
\begin{eqnarray}
  3 A_{i[jk]} \pi^{i'i}\varepsilon_{i'} + E^{i'}_{[j,k]}\varepsilon_{i'} + 
  E^{i'}_{j]}(\varepsilon_{i',[k} + \alpha_{i'[k}) = 0, \nonumber \\
  2 E^k_{[i,j]}\pi^{i'i}\varepsilon_{i'} - F^{i'k}_{,j}\varepsilon_{i'} + 
  E^{i'}_j (\varepsilon_{i',k'} + \alpha_{i'k'})\pi^{k'k} + C^{i'k}_j\varepsilon_{i'}) + 
  F^{i'k}(\varepsilon_{i',j} + \alpha_{i'j}) = 0,  \nonumber \\
  \frac{1}{2} F^{[jk]}_{,i}\pi^{i'i}\varepsilon_{i'} - 
  F^{i'[j}((\varepsilon_{i',k'} + \alpha_{i'k'})\pi^{k]k'} + C^{k]k'}_{i'}\varepsilon_{k'}) = 0, \nonumber \\
   E^j_i\pi^{i'i}\varepsilon_{i'} + F^{ii'}\varepsilon_{i'} = 0. \nonumber
\end{eqnarray}
Considering first the restrictions  from the first two of these 
equations coming from arbitrary symmetric $\alpha$'s 
and vanishing $\varepsilon$'s (they always belong to ${\cal GT}$),
one concludes that $E^i_j = a \cdot \delta^i_j$ and $F^{ij} = a \cdot \pi^{ij}$
for an arbitrary constant $a$. For this solution the forth equation is satisfied automatically 
 and the third one due to the twisted Jacobi identity. 
 
 Since the constructed superfunction is the extension of $H$ the values of
 all the components $A_{ijk}$ coincide with $\frac{1}{6}H_{ijk}$. It remains now 
 to fix the relative prefactor $a$ between $A$ and the remaining terms. 
 Due to the non-degeneracy condition that is asked in the theorem, 
 the first equation necessarily produces a restriction on $a$ which 
 is satisfied precisely for $a=1$, that is
 $$
  \tilde H = \frac{1}{6} H_{ijk} \tilde Q x^i \tilde Q x^j \tilde Q x^k
  + \frac{1}{2}\pi_{,i}^{jk}\tilde Q x^i p_j p_k + \tilde Q x^i \tilde Q p_i + \pi^{ij} \tilde Q p_i p_j,
$$
 Rewriting the expression in the usual basis of $T[1]T^*[1]M$
 completes the proof of the first point of the theorem.
 
 To verify the second point one needs simply to compute the 
 pull-back by $f^*$ of the resulting superfunction. Let us note here 
 that thanks to the choice of the basis on the fiber generated 
 by $\tilde Q$ the computation is very explicit using only the proposition 
 (\ref{prop:q-morphism}), this yields  
 \begin{eqnarray}
 S[X,A] = \int_{\Sigma_3} f^*(\tilde H) &=& \int_{\Sigma_3} \frac{1}{6} H_{ijk} \rd X^i \rd X^j \rd X^k+ \nonumber \\
  &+& \frac{1}{2}\pi_{,i}^{jk} \rd X^i A_j A_k + \rd X^i d A_i + \pi^{ij} \rd A_i A_j. \nonumber
\end{eqnarray}
 To recover precisely the functional of the twisted 
 Poisson sigma model one needs to perform the usual partial integration 
 of the result using Stokes' theorem. $\square$

\begin{rem}
If $H=0$, i.e. in the ``untwisted'' case, one obtains the functional of the 
Poisson sigma model up to a physically irrelevant constant prefactor.
One can also slightly generalize the picture for $H\neq0$, 
relaxing the condition that $\tilde H$ starts with $H$.
Then in the resulting formula for $\tilde H$ one can modify $A=H$ by a term
 $H'$ such that it is still closed and the contraction of $H'$ with the  anchor map $\pi^\#$ vanishes.
\end{rem}

\begin{rem} We should also note, that in the pure (untwisted) Poisson case one can ignore 
the $2$-tensor contribution of 
$\bar \alpha$ in (\ref{eps}) and restrict oneself to the analysis of $T^*[1]M$ as the manifold 
where the generators of the gauge transformations are defined (i.e. completely reverse the statement of the proposition \ref{th:sym}). 
However in the twisted case one does have to consider the $2$-tensor contribution, as for non-zero 
$H$ (\ref{magic_big}) could be very restrictive on $\varepsilon$, that is the Lie algebra 
${\cal G}$ would be too small to define uniquely $\tilde H$.
\end{rem}

\begin{rem}
  The condition of non-degeneracy of the pull-back of $H$ to the orbits of $\Pi$
  seems restrictive from the first sight. But actually this is a sufficient 
  condition that permits to guarantee a relation between the two parts of $\tilde H$
  not describing explicitly the subset of ${\cal GT}$ generated by $\varepsilon$'s for vanishing 
  $\alpha$'s.
  We know however some examples when the condition is not fulfilled but the extension 
  is still unique -- it would be interesting to study this phenomenon in details. 
\end{rem}

\begin{rem}
  The statement similar to the above theorem can be proved also for the Dirac sigma model (DSM) 
  (\cite{Kotov-Schaller-Strobl}, \cite{Salnikov-Strobl}), one can even attempt to generalize the construction 
  to a Lie algebroid structure on the target. 
  However the proof that we have presented here already encounters the key ideas and difficulties 
  and in contrast to the DSM we do not have to use any auxiliary structure. 
\end{rem}

\section{Equivariant cohomology of Courant algebroids} \label{sec:courant}
In this short section let us turn to a completely mathematical application of 
equivariant $Q$-cohomology. We will propose a definition of equivariant cohomology 
of Courant algebroids. We do it not with some computational purpose but more to 
show that the framework of the section \ref{sec:eqQcoh} is indeed large.

\begin{dfn}
A \emph{Courant algebroid} is a vector bundle $E \rightarrow M$ equipped with the following 
   operations: 
 a symmetric non-degenerate pairing $<\cdot , \cdot>$ on $E$, an $\mathbb{R}$-bilinear bracket 
 $[\cdot, \cdot]: \Gamma(E)\otimes \Gamma(E) \to \Gamma(E)$ on sections of $E$, and an anchor 
 $\rho$ which is a bundle map $\rho : E \to TM$, satisfying the axioms:
 \begin{eqnarray}
 \rho(\varphi) <\psi, \psi> =  2<[\varphi, \psi], \psi>, \nonumber \\
  \left[\varphi,[\psi_1, \psi_2]\right] = [[\varphi,\psi_1], \psi_2] + [\psi_1, [\varphi, \psi_2]], \label{leibniz}  \nonumber \\  
  2\left[\varphi, \varphi \right] = 
 \rho^*(d<\varphi, \varphi>), \nonumber \label{cour_ax2}
 \end{eqnarray}
 where $\rho^* : T^*M \to E$ (identifying $E$ and $E^*$ by $<\cdot, \cdot>$). \newline
\end{dfn}
We will restrict the analysis to 
a \emph{twisted exact Courant algebroid structure}
on $E  = TM \oplus T^*M$, governed by a closed $3$-form $H$ on $M$ 
(for a review on the subject see for example \cite{Melchior-thesis}). 
The symmetric pairing is given by 
$  <v \oplus \eta, v' \oplus \eta'> = \eta(v') + \eta'(v), $
the anchor $ \rho(v \oplus \eta) = v$ and the bracket is 
the $H$-twisted Courant--Dorfman bracket
\begin{equation} \label{Dorfman}
   [v \oplus \eta, v' \oplus \eta'] =  
   [v,v']_{\text{Lie}} \oplus ({\cal L}_v \eta' - \iota_{v'} \rd \eta + \iota_v\iota_{v'} H).
\end{equation}
According to \cite{Roytenberg2002} 
Courant algebroids are in bijection with degree $2$ symplectic $Q$-manifolds, that is 
one can construct a $Q$ structure associated to $E$.
To do this we consider the graded manifold  ${\cal M} = T^*[2]T[1]M$ and choose 
local coordinates on it: $(p_i(1), \psi_i(2), \theta^i(1), x^i(0))$.
On $T^*[2]T[1]M$ there is a canonical (degree $-2$) Poisson bracket.
The $Q$-structure can thus be constructed as a hamiltonian vector field $Q_{CA} = \{ {\cal Q}, \cdot \}$, 
where ${\cal Q} = \psi_i \theta^i + \frac{1}{6}H_{ijk}\theta^i\theta^j\theta^k$.

For the degree $-1$ vector fields let us consider degree $+1$ super functions on 
${\cal M}$ and hamiltonian vector fields associated to them. The most general degree $+1$
function is of the form $\epsilon  = \alpha_i \theta^i + v^i p_i$ for $\alpha_i, v^i$ -- 
some smooth functions of $x$. Considering the $Q_{CA}$ derived bracket of the 
vector fields $ \varepsilon :=  \{\epsilon, \cdot\}$ one recovers the structure 
similar to (\ref{Dorfman}).

It is now natural to give the following definition:
\begin{dfn} \label{def_eqcoh}
  The \emph{equivariant cohomology of a Courant algebroid} $E$ is described by the definitions
  \ref{def:Ghor}, \ref{def:Geq}, \ref{def:Gbasic} with the  $Q$-structure $Q_{CA}$ and
  the algebra ${\cal G} = \{\varepsilon\}$ or any subalgebra of it.
\end{dfn}
\begin{rem}
  Let us note that this definition already allows to treat the action of a huge algebra, 
  including for example all vector fields on $M$. But the construction is not limited to 
  it, namely we are not forced to restrict the vector fields $\varepsilon$ to hamiltonian ones, 
  instead we can consider all the degree $-1$ vector fields on ${\cal M}$.
  If even this is not enough we can perform the lift to $T[1]{\cal M}$ preserving the 
  pattern of the definition.
\end{rem}

\section{Conclusion/discussions}
In this paper we have observed on the example of the 
concept of equivariant $Q$-cohomology how the problem 
from theoretical physics can motivate the definition 
of a purely mathematical structure which can then again be used 
to analyze physical problems. Let us conclude by giving some 
small remarks and open questions inspired by the above presentation.

First, concerning the twisted Poisson sigma model:
we have obtained the condition (\ref{magic3}) describing the algebra 
of its symmetries (theorem \ref{th:sym}). 
It actually has a nice geometric interpretation: in the 
case of a non-degenerate closed $3$-form $H$ it is nothing but saying that 
$\pi^\# \varepsilon$ is a $2$-hamiltonian vector field with respect to 
a $2$-symplectic form $H$ (cf. \cite{Rogers}). This is even more visible 
in the analogous statement for the Dirac sigma model (\cite{Salnikov-Strobl}).
Moreover if one considers other dimensions of $\Sigma$ the condition
transforms to a system of partial differential equations including 
for example the stationary Lamb equation from hydrodynamics (\cite{Kozlov}). 
One is thus tempted to apply the information from these mathematical 
structures to study the symmetries of physical theories or conversely 
profit from knowledge of equivalent gauge theories to describe 
for example local geometry of twisted Poisson manifolds or Dirac structures.

Second, concerning the gauging problem: we have described a constructive procedure 
to produce gauge invariant functionals. We expect this approach to be 
useful to study some existing theories, like Lie algebroid Yang-Mills (\cite{LAYM})
or gauge theories on transitive Lie algebroids (\cite{morealgd}), 
as well as to produce new ones (like in the proof of the theorem \ref{prop:reverse}), since
there is no a-priori restriction neither on the dimension of the manifolds involved, 
nor on the gauge group. 
We have seen that the symmetry algebra  can be infinite dimensional, 
like the one defined by (\ref{magic3}), an interesting question to ask here 
is how to characterize the algebras for which gauging is not obstructed 
(\cite{Alekseev-Strobl}, \cite{Salnikov-Strobl2}).

Third, inspired by the construction of the section \ref{sec:courant} we expect the 
similar procedure to be related to existing definitions of equivariant cohomology 
of Lie algebroids (\cite{Roubtsov}, \cite{algd_coh}, \cite{algd_coh2}). \\[-0.2em]

\textbf{Acknowledgements.}
I would like to thank Thomas Strobl for constant attention towards this work. 
I also appreciate inspiring discussions with Alexei Kotov and Vladimir Roubtsov. 
This paper being an extended version of the poster presented at the FDIS13 
conference and the talk delivered during the CRM research program 
``Geometry and dynamics of integrable systems'', 
I would like to thank the organizers of these events for inviting me, as well as the 
participants for valuable comments and questions. 




\bibliographystyle{elsarticle-num}



\end{document}